\documentclass[aps,prl,twocolumn,showpacs,showkeys,10pt,superscriptaddress,preprintnumbers,nofootinbib]{revtex4-1}
\usepackage[T1]{fontenc} 
\usepackage{amssymb}
\usepackage{amsmath}
\usepackage{color}
\usepackage{xspace}
\usepackage{caption}
\usepackage{appendix}
\usepackage{float}
\usepackage{hyperref}
\usepackage{diagbox}
\usepackage{soul}
\usepackage{subfig}
\usepackage{slashed}
\usepackage{amsfonts}
\usepackage{multirow}
\usepackage{mathrsfs}
\usepackage{graphicx}
\usepackage{amsmath}
\usepackage{amssymb}
\usepackage{bm}
\usepackage{bbm}
\usepackage{color}

\newcommand{\nn}{\nonumber}
\newcommand{\bea}{\begin{align}}
\newcommand{\eea}{\end{align}}
\newcommand{\beq}{\begin{equation}}
\newcommand{\eeq}{\end{equation}}
\newcommand{\bqa}{\begin{eqnarray}}
\newcommand{\eqa}{\end{eqnarray}}

\allowdisplaybreaks[4]

\begin{document}

\title{\large\bf Analytic result for the top-quark width at next-to-next-to-leading order in QCD}

\author{Long-Bin Chen}
\affiliation{School of Physics and Materials Science, Guangzhou University, Guangzhou 510006, China}

\author{Hai Tao Li}
\affiliation{School of Physics, Shandong University, Jinan, Shandong 250100, China}

\author{Jian Wang}
\email{j.wang@sdu.edu.cn}
\affiliation{School of Physics, Shandong University, Jinan, Shandong 250100, China}

\author{Yefan Wang}
\email{wangyefan@sdu.edu.cn}
\affiliation{School of Physics, Shandong University, Jinan, Shandong 250100, China}

\begin{abstract}
We present the first full analytic results of  next-to-next-to-leading order (NNLO) QCD corrections to the top-quark decay width $\Gamma(t\to  Wb)$ by calculating the imaginary part of three-loop top-quark self-energy diagrams.
The results are all expressed in terms of harmonic polylogarithms and valid in the whole region $0\le m_W^2\le m_t^2$.
The expansions in the $m_W^2\to 0$ and $m_W^2\to m_t^2$ limits coincide with previous studies.
Our results  can also be taken as the exact prediction for the lepton invariant mass spectrum in semileptonic $b\to u$ decays.
We also analytically compute the decay width including the off-shell $W$ boson effect up to NNLO in QCD for the first time. 
Combining these contributions with electroweak corrections and the finite $b$-quark mass effect, we determine  the most precise top-quark width to be 1.331 GeV for $m_t=172.69$ GeV.
The total theoretical uncertainties including those from renormalization scale choice, top-quark mass renormalization scheme, input parameters and missing higher-order corrections   are scrutinized and found to be less than $1\%$.
\end{abstract}

\keywords{Top quark, Decay width, QCD corrections}

\maketitle

\section{ Introduction}
As the heaviest elementary particle in the standard model (SM) of particle physics, the top quark plays an important role in studies of fundamental interactions. 
Its properties have been investigated in great detail
since its discovery at the Tevatron \cite{D0:1995jca,CDF:1995wbb}.
Among them, the top-quark decay width $\Gamma_t$ is one of the most important parameters.
The large value of this quantity indicates that the top quark has a lifetime much shorter than the period for its hadronization \cite{Bigi:1986jk}. 
Thus, we can measure directly the properties of the top quark itself, rather than the hadrons formed by top quarks. 
Such  studies  on top quarks provide an excellent playground for 
the precision test of the SM and the search for new physics signals.

The top quark can be produced  via both strong and electroweak interactions, but  decays only by electroweak interaction.
In the SM, it decays almost exclusively to $Wb$ and therefore its decay width is determined by this decay mode, i.e., $\Gamma_t=\Gamma(t\to Wb)$.

The top-quark decay width can be measured in various ways.
In the first method, one could compare the shape of the reconstructed mass distribution of top quarks with samples in which the top-quark width is already known.
This method relies on detector resolution and precise calibration of the jet energy scale.
The missing momentum of the neutrino from top quark decay causes large uncertainties in the determination of the width.
The ATLAS collaboration has  measured the width to be 
$\Gamma_t=1.9 \pm 0.5$ GeV using this method \cite{ATLAS:2019onj}.

One may also take an indirect approach by combining the information of the branching fraction ratio $B(t\to Wb)/B(t\to Wq)$ from top-quark pair production and that of the  t-channel single top-quark cross section. 
The CMS collaboration has performed such a measurement and determined the top-quark total decay width $\Gamma_t = 1.36 \pm  0.02~(\rm stat.)^{+0.14}_{-0.11}~(\rm syst.)$ GeV \cite{CMS:2014mxl}.

Novel methods have been proposed recently.
The top-quark decay width can be directly probed by measuring the on-/off-shell ratio of $b$-charge asymmetry from $pp\to bWj$,
and a $0.2$-$0.3$ GeV  precision is expected at the high luminosity LHC \cite{Giardino:2017hva}.
Applying the same idea to the top-quark pair production, the top-quark width can be constrained with an uncertainty of $12\%$ assuming an experimental accuracy of $5\%$
\cite{Baskakov:2018huw}.
A more realistic analysis of the ATLAS differential cross section measurement shows that a result of $1.28 \pm 0.31$ GeV for the width can be obtained
\cite{Herwig:2019obz}.

The top-quark width can also be measured at a future $e^+ e^-$ collider. 
Following a multi-parameter fit approach, it can be extracted with an uncertainty of $30$ MeV \cite{Martinez:2002st}.
The sensitivity would be further improved by using polarized beams or an optimized scan strategy, and an accuracy  of $21$ (26) MeV can be obtained at the ILC \cite{Horiguchi:2013wra} (CEPC \cite{Li:2022iav}).
The  measurement at the CLIC will be at the same level \cite{CLICdp:2018esa}.

On the theoretical side, 
the next-to-leading order (NLO) quantum chromodynamics (QCD) corrections to the $t\to Wb$ decay were  first computed analytically in  Refs.~\cite{Jezabek:1988iv,Czarnecki:1990kv,Li:1990qf} before the discovery of the top quark.
The NLO electroweak (EW) corrections were presented around the same time \cite{Denner:1990ns,Eilam:1991iz}.
The next-to-next-to-leading order (NNLO) QCD corrections to the top-quark decay width have been calculated in Refs. \cite{Czarnecki:1998qc,Chetyrkin:1999ju,Blokland:2004ye,Blokland:2005vq} 
and Ref. \cite{Czarnecki:2001cz} about twenty years ago using asymptotic expansion in the $w \equiv  m_W^2/m_t^2\to 0$ and $w\to 1$ limit, respectively.
Later, the total and differential decay widths were calculated numerically~\cite{Gao:2012ja,Brucherseifer:2013iv},
and polarized decay rates were also studied \cite{Czarnecki:2010gb,Czarnecki:2018vwh}. 
Recently, the renormalization scheme and scale uncertainties in this process were discussed in \cite{Meng:2022htg}.
However, the full analytic results of NNLO QCD corrections valid for any $w$ from 0 to 1 are still unknown.
They are helpful not only in understanding the mathematical structure of the scattering amplitude at the multiple-loop level but also  in providing fast and accurate numerical results for phenomenological analyses.

\section{Calculation methods}
We calculate the top-quark decay width $ \Gamma_t$ by using the optical theorem to relate it to the imaginary part of top quark self-energy diagrams $\Sigma$ for the process $t\to Wb\to t$,
\beq
\Gamma_t=\frac{\text{Im}(\Sigma)}{m_t}\, .
\eeq
In this way, we are not bothered by the divergences that exist in the virtual and real corrections separately and the complicated phase space integration.

\begin{figure}
\includegraphics[width=1.0\linewidth]{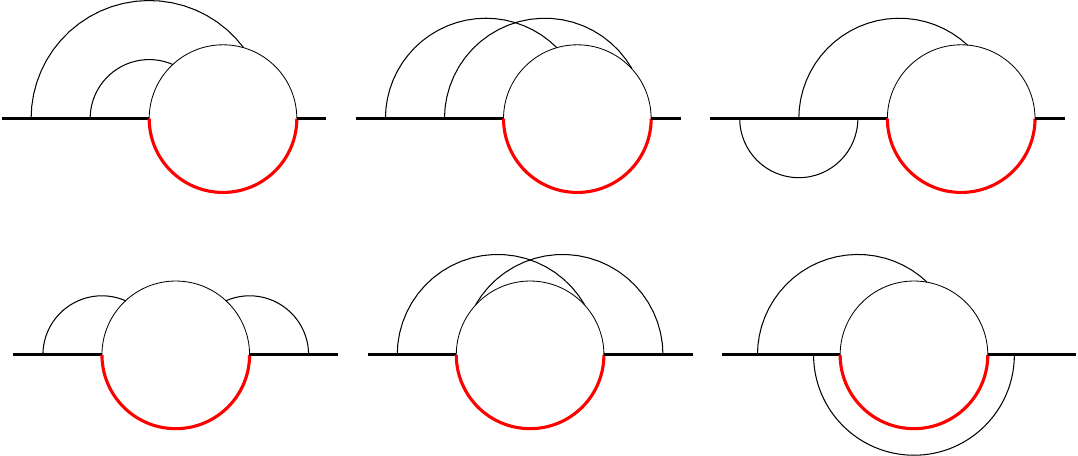}
    \caption{The topologies of three-loop Feynman diagrams  that contribute to the top-quark decay  $t\to b W$ at NNLO in QCD. The thick and thin black lines stand for the top quarks and massless particles, respectively. The red lines represent the $W$ bosons.}
    \label{fdiag}
\end{figure}

The amplitudes of the self-energy diagrams $\Sigma$ can be  generated by using the packages {\tt FeynArts} \cite{Hahn:2000kx} and {\tt FeynCalc}~\cite{Shtabovenko:2020gxv}. 
Due to the angular momentum conservation, the final- and initial-state top quarks have the same spin.
Performing summation over all the spins of the external top quarks,  each amplitude is converted to a trace along the fermion line,
and thus consists of  scalar loop integrals.
They are reduced to a minimal set of integrals called master integrals (MIs) using the identities induced from integration by parts  \cite{Tkachov:1981wb,Chetyrkin:1981qh}.
In this step, we have made use of the package {\tt FIRE}~\cite{Smirnov:2019qkx}.
There are nine integral families at the three-loop level.
After reduction, however, we are left with six kinds of topologies, as shown in Fig.~\ref{fdiag}.

The imaginary part of $\Sigma$ receives contributions from cut diagrams where some of the internal propagators can be put on-shell simultaneously \cite{doi:10.1063/1.1703676}. 
In particular, the $W$ boson propagator and at least one $b$ quark propagator should be cut. 
The MIs containing such cuts are labeled  cut MIs.

Then we construct canonical differential equations for the cut MIs by choosing a proper basis $\textbf{I}(w,\epsilon)$ such that the dimensional regularisation parameter $\epsilon=(4-d)/2$, where $d$ is the space-time dimension, decouples from the kinematic variables \cite{Kotikov:1990kg,Kotikov:1991pm,Henn:2013pwa},
\beq 
\text{d}\, \textbf{I}(w,\epsilon) = \epsilon\, \text{d} \left[\sum_{i=1}^4 \textbf{R}_i \log(l_i)\right] \textbf{I}(w,\epsilon)
\label{eform}
\eeq
with the letters $l_i\in\{w-2,w-1,w,w+1\}$ and $\textbf{R}_i$ being rational matrices. 
The explicit form of the canonical basis $\textbf{I}$ and the  differential equations  in Eq.~(\ref{eform}) are available upon request from the authors.
It is highly nontrivial to achieve the canonical form for a three-loop integral basis that contains two different masses in the propagators.

In order to solve the above differential equations, 
boundary conditions have to be provided.
Most of the basis integrals are regular at $w=0$, i.e., they do not contain any logarithmic  structure $\log(w)$, and thus can be obtained from the results for  heavy-to-light decay processes \cite{vanRitbergen:1999fi,Blokland:2005vq}, or by using the regularity conditions of the differential equations at $w=0$.
The calculation of the basis integrals that are not regular is more technical.
Some of them can be calculated directly.
The others can be determined up to a constant after solving the differential equations.
This constant is firstly computed numerically with an over 50-digit accuracy employing the {\tt AMFlow} package \cite{Liu:2017jxz,Liu:2022chg}, and then reconstructed in analytic form using the PSLQ algorithm \cite{Ferguson1992,Ferguson:1999aa}.

The results of the basis integrals are all expressed in terms of multiple polylogarithms \cite{Goncharov:1998kja} with arguments $l_i$.
In particular, we find that  the letter $l_1=w-2$  appears  always along with $w-1$ and $w$.
Therefore, we can  change the variable  $w\to 1-w$ in those integrals containing the letter $w-2$, making all the analytic results of the master integrals   written simply in terms of  harmonic polylogarithms (HPLs) as defined in \cite{Remiddi:1999ew}.

Combining the cut MIs and their corresponding coefficients, we obtain  analytical results for the imaginary part of  three-loop top-quark self-energy diagrams,
which are free of infrared divergences but still ultraviolet divergent.
We then calculate the contribution of the counter-terms following the standard renormalization procedure,
and find that the ultraviolet divergences cancel out exactly.

\section{Analytical and Numerical Results}
The top-quark decay width for $t\to Wb $ at NNLO in QCD can be expressed as 
\bqa
&&\Gamma(t\to  Wb)=\Gamma_0\left[X_0+\frac{\alpha_s}{\pi}X_1+\left(\frac{\alpha_s}{\pi}\right)^2 X_2\right],
\eqa
where  $\Gamma_0=\frac{G_F m_t^3|V_{tb}|^2}{8\sqrt{2}\pi} $  and the coefficients at each order of the strong coupling $\alpha_s$ read
\bqa
X_0 &=& (2w+1)(w-1)^2, \nonumber\\
X_1 &=&C_F\bigg( X_0\Big(-2H_{0,1}(w)+H_{0}(w)H_{1}(w)-\frac{\pi^2}{3}\Big)\nonumber\\
&+&\frac{1}{2}(4w+5)(w-1)^2H_1(w)\nonumber\\
&+&w(2w^2+w-1)H_0(w) \nonumber \\
&+&\frac{1}{4}(6w^3-15w^2+4w+5)\bigg),\\
X_2 &=& C_F(T_R n_l X_l + T_R n_h X_h + C_F X_F +C_A X_A).\nonumber
\eqa
Here we have taken a massless $b$ quark for simplicity. 
The results for $X_0$ and $X_1$ have been well-known \cite{Jezabek:1988iv,Czarnecki:1990kv,Li:1990qf}.
The full analytic form of $X_2$ is new and constitutes one of the main results of the present work.
It has been decomposed in gauge-invariant color structures, which 
 are specified in QCD by $C_F=4/3,C_A=3,T_R=1/2$,
and $n_l~(n_h)$  the number of massless (massive) quark species.
The coefficients of each color structure at the renormalization scale $\mu=m_t$ are given by
\begin{widetext}
\begin{eqnarray}
X_l&=&
-\frac{X_0}{3} \left[H_{0,1,0}(w)-H_{0,0,1}(w)-2 H_{0,1,1}(w)+2 H_{1,1,0}(w)-\pi ^2 H_1(w)-3 \zeta (3)\right]+ g_l(w),\nonumber\\
X_h &=& -\frac{\left(X_0-12 w\right)}{3} \left[\zeta (3)-H_{0,0,1}(w)\right]+g_h(w),\nonumber\\
X_F&=&\frac{1}{12} X_0 \big[-6 \left(2 H_{0,1,0,1}(w)+6 H_{1,0,0,1}(w)-3 H_{1,0,1,0}(w)-12 \zeta (3) H_1(w)\right)
-\pi ^2 H_{1,0}(w)\big]\nonumber\\&+&\left(X_0+4 w\right) \left(-\frac{1}{6} \pi ^2 H_{0,-1}(w)-2 H_{0,-1,0,1}(w)\right)\nonumber\\
&+&\frac{1}{12}  \left(18 w^3-3 w^2+76 w+15\right)\pi ^2 H_{0,1}(w)-\frac{1}{2} \left(4 w^3-2 w^2+4 w+3\right) H_{0,0,0,1}(w)\nonumber\\
&+&\frac{1}{2} \left(4 w^3-2 w^2+16
 w+3\right) H_{0,0,1,0}(w)+w \left(2 w^2-7 w-16\right) H_{0,0,1,1}(w)\nonumber\\
&-&\frac{1}{2} \left(2 w^3-11 w^2-28 w-1\right) H_{0,1,1,0}(w)+\frac{1}{720} \pi ^4 \left(42 w^3-191 w^2-328 w-11\right)
+ g_F(w),\nonumber\\
X_A&=&\frac{1}{8} X_0 \left[-\pi ^2 H_{1,0}(w)+8 H_{1,0,0,1}(w)-2 H_{1,0,1,0}(w)-12\zeta(3)H_1(w)\right]\nonumber\\
&+&\frac{1}{24}  \left(10 w^3+33 w^2+44 w+11\right)\pi ^2 H_{0,1}(w)-\frac{1}{4} \left(8 w^2+16 w+1\right) H_{0,0,0,1}(w)\nonumber\\
&+&\left(X_0+4 w\right) \left(\frac{1}{12} \pi ^2 H_{0,-1}(w)+H_{0,-1,0,1}(w)\right)+\frac{1 }{1440}\pi ^4\left(86 w^3-385 w^2-312 w+11\right)\nonumber\\
&-&\frac{1}{4} \left(8 w^2+4 w+1\right) \left(2 H_{0,0,1,1}(w)-H_{0,0,1,0}(w)\right)+\frac{1}{4} \left(2 w^3+13 w^2+12 w+3\right) H_{0,1,1,0}(w) 
+ g_A(w).
\label{eq:Xeq}
\end{eqnarray}
\end{widetext}
Here we have shown explicitly in each coefficient the results of maximal transcendental {\it weight} of HPLs, which is defined by the dimension of the vector $\vec{m}$ in $H_{\vec{m}}(w)$. 
The transcendental weight can be added in each term and
the ubiquitous constants $\pi$ and Riemann Zeta function $\zeta(n)$ should be considered of weight one and $n$, respectively, since $\pi = -i H_0(-1+i0)$ and $\zeta(n) = H_{\vec{0}_{n-1},1}(1)$. 
The results of lower transcendental weight are denoted by the functions $g_l(w),g_h(w), g_F(w)$, and $g_A(w)$, of which the explicit forms can be found in the appendix.
The above results are obtained at the renormalization scale $\mu=m_t$.
The scale-dependent part in  these coefficients can be recovered using the fact the total decay width is scale-independent.

We have made multiple checks at various stages of the calculation.
All the analytic results for the master integrals  have been confirmed against the numerical {\tt AMFlow} package  \cite{Liu:2022chg}.
The amplitudes have been calculated in two different gauges for the $W$ boson propagator, 
i.e., the Feynman-'t Hooft gauge and the unitary gauge, and perfect agreement is found.
The particles appearing in the loops and the renormalization constants are all different,
and thus the agreement between the results in these two gauges provides a strong check of the correctness of our calculation.
Our analytic results can  be expanded to any fixed order around a  $w$ from 0 to 1.
The expansion up to $\mathcal{O}(w^5)$ coincides with the result reported in \cite{Blokland:2004ye,Blokland:2005vq},
and the expansion around  $w=1$ reproduces the asymptotic expansion result given in \cite{Czarnecki:2001cz}.

Though the decay width at $w=0$ and $w=1$ is finite, it exhibits logarithmic structures near these boundaries.
We have extracted such logarithmic terms by making use of the shuffle algebra properties  of the HPLs.
They are shown in the appendix.
It would be interesting if these logarithms could be reproduced  and resummed to all orders from effective field theory.

\begin{figure}[ht]
    \includegraphics[width=0.8\linewidth]{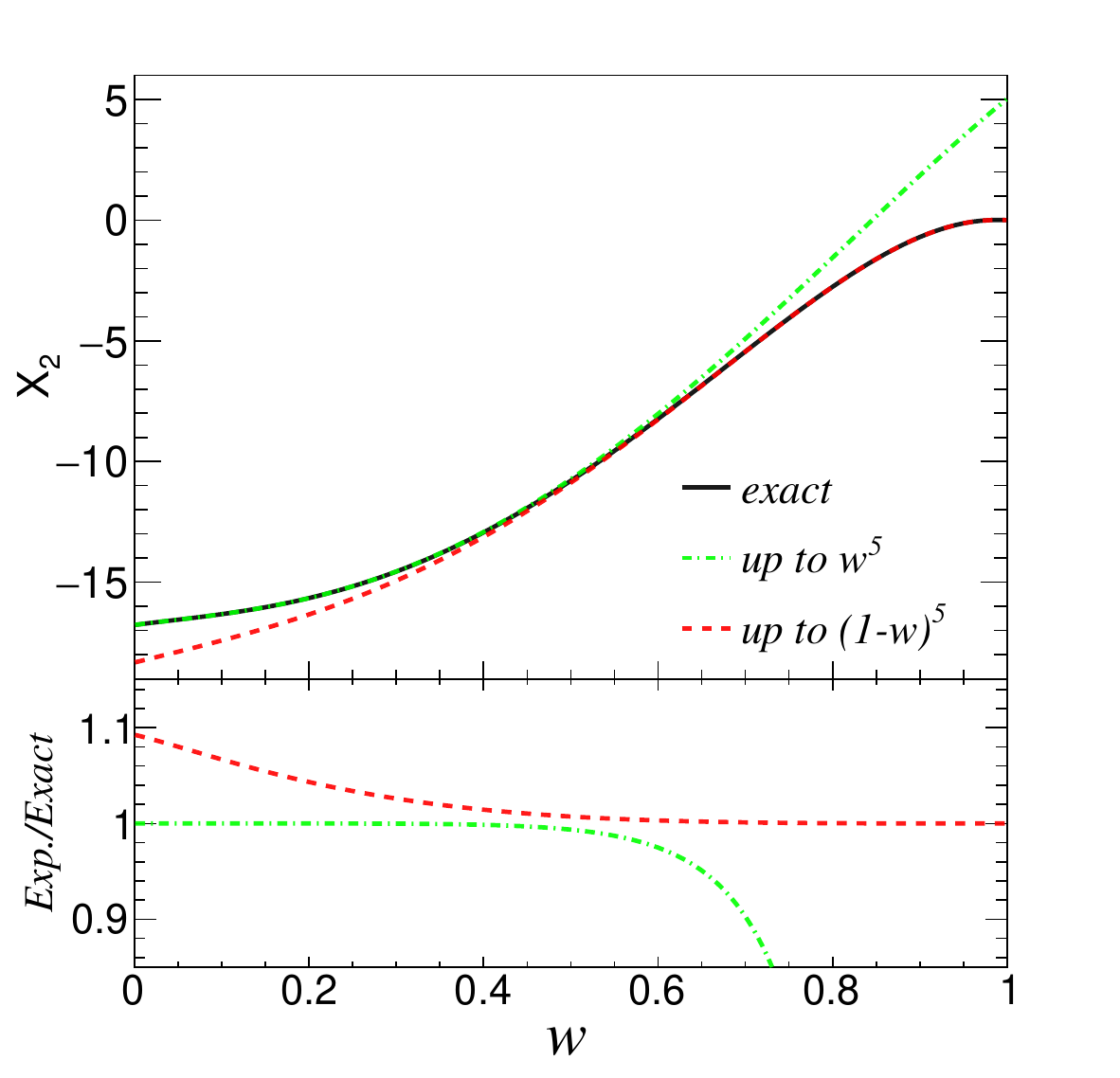}
    \caption{Comparison between the exact result of $X_2$ (black line) with its expansion around $w=0$ (dot-dashed green lines) or $w=1$ (dashed red lines).
    The curves in the lower panel represent the deviation of the expansions from the exact result.
    }
    \label{fig:myto1_label}
\end{figure}

To illustrate the difference between our exact results and the approximations in series expansions, we show the numerical values in Fig.~\ref{fig:myto1_label}.
The expansion up to order $w^5$,
which was given explicitly in Refs. \cite{Blokland:2004ye,Blokland:2005vq},
agrees with the exact result very well for $w<0.5$, but begins to deviate from it for larger $w$.
The expansion value near $w=1$ is not approaching zero and is thus irrational, since the phase space is  nearly prohibited in this region. 
At the other end, the difference between the series up to $(1-w)^5$, which was collected in Ref. \cite{Blokland:2005vq}, and the exact result is negligible for $w>0.5$, but becomes sizable when $w$ decreases.
The most obvious deviation is about $10\%$ at $w=0$.
Our analytic results unify the two  
expansions and are valid in the entire interval $0\le w \le 1$.

The top quark decay is closely related to other important processes.
Our results multiplied by a constant factor can be taken as the exact prediction  for the lepton invariant mass spectrum in semileptonic $b\to u$ decays.
After integration over $w$ from 0 to 1 analytically, we reproduce the NNLO QCD correction to the total decay rate of $b$ quark semileptonic decay  $\Gamma(b\to X_u e \bar{\nu}_e)$ \cite{vanRitbergen:1999gs}, which is a requisite to extract a precise value of the   Cabibbo-Kobayashi-Maskawa matrix element $V_{ub}$ from $B$ meson experiments.
Furthermore, if we integrate only the abelian contribution $X_F$ over $w$, we obtain the analytic two-loop QED correction to the muon lifetime \cite{vanRitbergen:1998yd}, which has been used to derive an accurate value for the Fermi coupling constant
$G_F$ \cite{MuLan:2012sih}.

In the above discussion, the $W$ boson in the top quark decay is assumed  on the mass shell.
In reality, it has a width of $\Gamma_W= 2.085 $ GeV~\cite{ParticleDataGroup:2022pth}
since it could  decay immediately into leptons or quarks.
After considering the fact that the top quark can decay into an off-shell $W^*$ boson, the top-quark decay width is given by \cite{Jezabek:1988iv}
\begin{align} \label{eq:gammaW}
   \tilde{\Gamma}_t & \equiv   \Gamma(t\to W^* b ) \nn \\
  & = \frac{1}{\pi}\int_0^{m_t^2}  dq^2 \frac{m_W \Gamma_W }{(q^2-m_W^2)^2 + m_W^2 \Gamma_W^2 } \Gamma_t(q^2/m_t^2).
\end{align}
With the analytical result of $\Gamma_t(x)$ at hand, it is straightforward to perform the integration and obtain the analytical form of $\tilde{\Gamma}_t$ in terms of multiple polylogarithms \cite{Goncharov:1998kja}, which is provided in the appendix. 
This is another new result of our work.

\begin{table}[]
    \centering
    \begin{tabular}{c|ccccc}
    \hline \hline 
            & $\delta_b^{(i)}$  & $\delta_{ W}^{(i)}$ & $\delta_{\rm EW}^{(i)}$ & $\delta_{\rm QCD}^{(i)}$& $\Gamma_t$ [GeV]   \\
    \hline 
        LO    & -0.273  & -1.544   &  $-$ & $-$    &  1.459  \\
        NLO   & 0.126   & 0.132     &  1.683    & -8.575 &  1.361 \\
        NNLO  & $*$     & 0.030    & $*$  & -2.070 &  1.331 \\
    \hline \hline 
    \end{tabular}
    \caption{Top-quark width up to NNLO and corrections  in percentage ($\%$) from  finite $b$-quark mass effect, off-shell $W$ boson contribution, EW  and QCD higher orders normalized by the LO width $\Gamma_{t}^{(0)} = 1.486$ GeV with $m_b=0$ and on-shell $W$. 
    The values of $\delta_b^{(1)}$ and $\delta_{\rm EW}^{(1)}$ are obtained from the formulae given in \cite{Bohm:1986rj,Jezabek:1988iv,Denner:1990ns,Denner:1990tx}.
    The symbol `$*$' denotes the contribution that has not been calculated yet.
    The last column gives the decay width including all the possible corrections up to that order.}
    \label{tab:my_label}
\end{table}

Now we provide numerical results for the top-quark decay width \footnote{
All the above formulae are incorporated in a  Mathematica program {\tt TopWidth} 
which can be downloaded from  \url{https://github.com/haitaoli1/TopWidth}.
The program has been organized in a form that can easily be used. }.
The input parameters are given by~\cite{ParticleDataGroup:2022pth}
\begin{align}
    m_t & =172.69 ~{\rm GeV },\quad
     m_b=4.78~{\rm GeV },  \nn \\
   m_W & =80.377 ~{\rm GeV },\quad \Gamma_W=2.085~{\rm GeV }, \\
   m_Z&= 91.1876 ~{\rm GeV}, \quad G_F  = 1.16638 \times 10^{-5}~{\rm GeV }^{-2}. \nn
\label{eq:input}    
\end{align}
We choose the Cabibbo-Kobayashi-Maskawa matrix element  $|V_{tb}| =1 $,
and  $\alpha_s(m_Z) =0.1179$.
The values of $\alpha_s$ at other scales are derived using the three-loop renormalization group evolution equation~\cite{Gardi:1998qr,Deur:2016tte}.
The decay width is decomposed according to the perturbative orders,
\begin{align}
    \Gamma_t & = \Gamma_t^{(0)} [(1+\delta_b^{(0)} + \delta_W^{(0)}) \nn \\
    & \qquad + (\delta_b^{(1)} + \delta_W^{(1)}  +\delta_{\rm EW}^{(1)} +\delta_{\rm QCD}^{(1)}  ) \\ 
    & \qquad + (\delta_b^{(2)} + \delta_W^{(2)}  +\delta_{\rm EW}^{(2)} +\delta_{\rm QCD}^{(2)}  +\delta_{\rm EW \times QCD}^{(2)}   ) ],\nn
\end{align}
where the LO width $\Gamma_{t}^{(0)} = 1.486$ GeV with $m_b=0$ and $W$ on-shell, 
$\delta_b^{(i)}$ and $\delta_W^{(i)}$ denote the corrections from finite $b$ quark mass
effect and off-shell $W$ boson contribution, respectively.
The higher-order EW and QCD corrections are indicated by $\delta_{\rm EW}^{(i)}$ and $\delta_{\rm QCD}^{(i)}$, respectively.
The superscripts specify the perturbative order in which they contribute.

In Table \ref{tab:my_label} we show their individual contributions.
We can see that the dominant corrections come from QCD higher orders, which are $-8.58\%$ and $-2.07\%$ at NLO and NNLO, respectively.
The NLO EW correction,  calculated using the analytic expressions in \cite{Denner:1990ns,Denner:1990tx}, increases the LO result by $1.68\%$.
The off-shell $W$ boson contributes a $-1.54\%$ correction at LO,
while its effect is only $0.13\%$ at NLO, nearly amounting to $\delta_{W}^{(1)}\times \delta_{\rm QCD}^{(1)}$.
The  off-shell $W$ boson effect at NNLO is further suppressed.
The $b$ quark mass correction at LO is $-0.27\%$, as expected at the same order of $m_b^2/m_t^2$.
The modification at NLO is not severely suppressed compared to the LO one.
We have checked that this is due to the  large logarithms at subleading power. 
It would be interesting to investigate their structure following the method in \cite{Beneke:2022obx}.

\begin{figure}[ht]
    \includegraphics[width=0.9\linewidth]{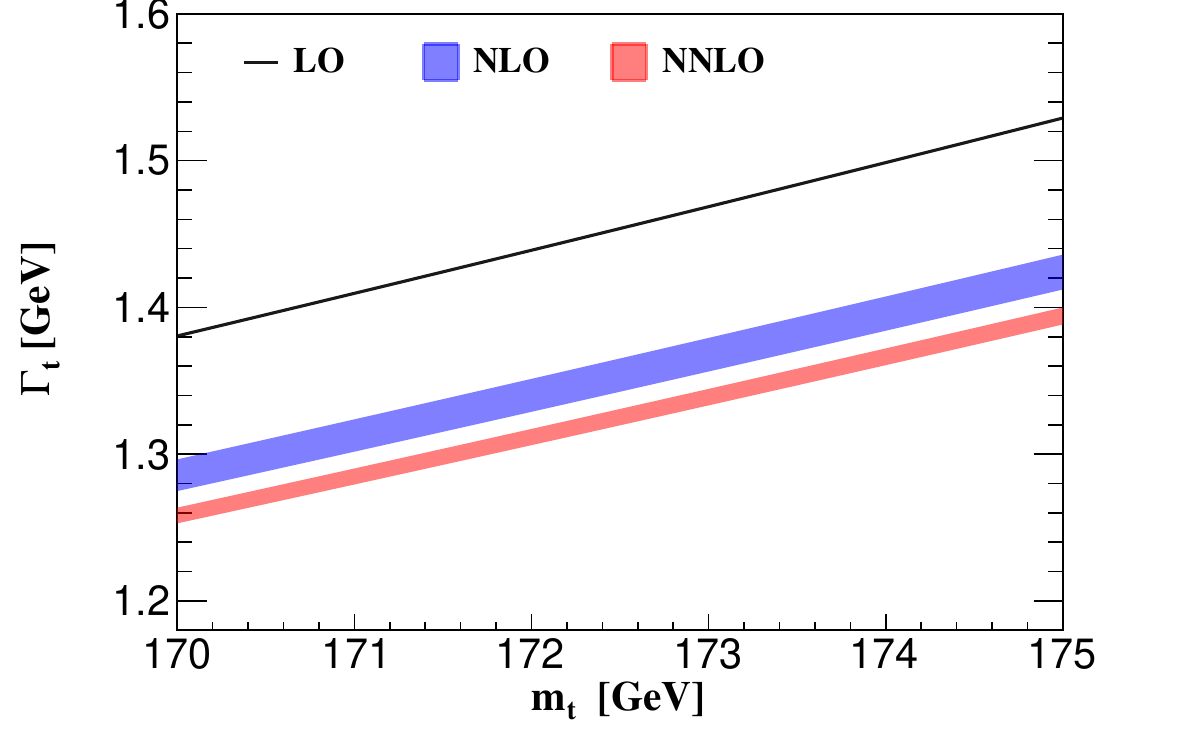}
    \caption{ Top-quark width as a function of $m_t$.  The  bands denote the QCD scale uncertainties.
    }
    \label{fig:GammaT}
\end{figure}

Collecting all the contributions as shown in Table \ref{tab:my_label}, we obtain the  top-quark width  $\Gamma_t=1.331$ GeV,
which is the most precise determination of this quantity to date. 
When the top-quark mass varies from 170 GeV to 175 GeV, the width changes from 1.258 GeV to 1.394 GeV, displaying an almost linear dependence within this range, as shown in Fig.~\ref{fig:GammaT}.

Finally, we discuss the theoretical uncertainties in our results.
The first uncertainty is due to the arbitrary choice of the QCD renormalization scale $\mu$.
We have chosen the default value $\mu=m_t$ in the numerical evaluation. 
Now we scan the scale $\mu \in [m_t/2,2m_t]$ and find that the variation of the result is about   $\pm 0.8\%$ and $\pm 0.4\%$ at NLO and NNLO, respectively, which can be seen in Fig.~\ref{fig:GammaT}.
This scale uncertainty has been reduced dramatically after including NNLO QCD corrections. 
The second uncertainty comes from the renormalization scheme of the top-quark mass.
The top-quark mass used in Eq. (\ref{eq:input}) is defined in the on-shell renormalization scheme.
If we adopt the $\overline{\rm MS}$  scheme  in QCD corrections, the top-quark decay width would be $1.309$ GeV at NLO and $1.332$ GeV at NNLO,
which differ from the results using the on-shell scheme by $- 3.79\%$ and $ 0.09\%$ at NLO and NNLO, respectively.
The perturbative series using the on-shell mass usually grows rapidly at higher orders due to the infrared renormalon divergence \cite{Beneke:2021lkq}.
This problem can be avoided if the decay rate is expressed in terms of the $\overline{\rm MS}$ renormalized top-quark mass. Assuming a power-like growth for the coefficients of $(\alpha_s/\pi)^n$ \cite{vanRitbergen:1999gs}, the missing NNNLO QCD contribution would be of the order of $0.4\%$.
This is consistent with the expectation that the scale dependence gives a rough estimate of the unknown higher-order contributions.  
Third,  the uncertainties at NNLO from the input parameters $\alpha_s (m_Z) = 0.1179\pm 0.0009  $ and $ m_W = 80.377\pm 0.012 $ GeV \cite{ParticleDataGroup:2022pth} are $0.1\%$ and $0.01\%$, respectively. 
Fourth, the deviation between the $\alpha$ scheme and the $G_F$ scheme we have used in the EW correction is $0.1\%$ at NLO.
Lastly, the NNLO EW as well as the mixed ${\rm EW \times QCD}$ corrections have not been studied so far, but we estimate that they are of the order of $\alpha\delta_{\rm EW}^{(1)}$ and $\delta_{\rm EW}^{(1)} \times \delta_{\rm QCD}^{(1)}$, respectively.
Therefore, after considering all the possible uncertainties, we conclude that the uncertainty of our result at NNLO is less than $1\%$.

\section{Conclusion}
We have provided the first full analytical result of the top-quark width at NNLO in QCD.  
The  result is obtained using the optical theorem and expressed in terms of only harmonic polylogarithms, and thus it enables a fast and exact evaluation.
The off-shell $W$ boson contribution is also calculated analytically up to NNLO in QCD for the first time.
Combining our results with NLO EW corrections and  finite $b$ quark mass effects, 
the most precise top-quark width is predicted to be 1.331 GeV for $m_t=172.69$ GeV with the total theoretical uncertainty less than $1 \%$.

\section*{Acknowledgements}
We would like to thank Jun Gao, Zhao Li, and Yu-Ming Wang for discussions.
We thank Ansgar Denner for the communication on the implementation of the EW corrections.
This work was supported in part by the National Natural Science Foundation of China under grant  Nos. 12005117, 12147154, 12175048, 12275156.  
The work of L.B.C. was also supported by the Natural Science Foundation of Guangdong Province under grant No. 2022A1515010041. The work of J.W. was also supported by the Taishan Scholar Project of Shandong Province (tsqn201909011).

\newpage

\onecolumngrid
\newpage

\appendix
\section*{Appendix}
\label{sec:appendix}

The functions $g_l(w),g_h(w), g_F(w)$ and $g_A(w)$ in Eq.(\ref{eq:Xeq}) are given by 
\bqa
g_l(w) &=& -\frac{\left(38 w^2-55 w-37\right) (w-1) H_{1,0}(w)}{18 }+\frac{\left(7 w^3-39 w^2+15 w+5\right) H_{0,1}(w)}{9 }\nonumber\\
&-&\frac{(4w+5)(w-1)^2 H_{1,1}(w)}{3 }+\frac{\pi ^2 \left(124 w^3-111 w^2-12 w+23\right)}{108 }\nonumber\\
&-&\frac{\left(124 w^3-35 w^2-143 w+6\right) (w-1) H_1(w)}{36 w}
-\frac{1}{36} w\left(106 w^2-25 w-86\right) H_0(w)\nonumber\\
&-&\frac{\left(99 w^2-120 w-22\right) (w-1)}{36 },\nonumber\\
g_h(w) &=&-\frac{\pi ^2 (w-1) \left(11 w^2-13 w-10\right)}{18 }+\frac{\left(19 w^4+32 w^3-18 w^2-8 w+23\right) \left(\frac{\pi ^2}{6}-H_{0,1}(w)\right)}{9 (w-1)}\nonumber\\
&+&\frac{\left(265 w^4+168 w^3-498 w^2+344 w+9\right) H_1(w)}{54w}
+\frac{15902 w^3-9237 w^2-12528 w+12775}{1296 },\nonumber\\
g_F(w) &=&-\frac{1}{96} \left(w^2-12\right) \big(24 H_{-1,0,0}(1-w)+24 H_{-1,0,1}(1-w)+14 \pi ^2 H_{-1}(1-w)-3 \zeta (3)\nonumber\\&-&18 \pi ^2 \log (2)\big)+\left(5 w^2+8 w+3\right) (w-1) \left(2 H_{-1,0,1}(w)+\frac{1}{6} \pi^2 H_{-1}(w)\right)\nonumber\\
&+&\frac{1}{2} \left(w^2-25 w-26\right) (w-1) H_{1,1,0}(w)-\frac{1}{2} \left(18 w^3-9 w^2-4 w+3\right) H_{0,0,1}(w)\nonumber\\
&+&\frac{1}{4} \left(2 w^3-15w^2+10w+12\right)H_{0,1,0}(w)+\frac{1}{4} \left(4 w^3+57 w^2+4 w-54\right) H_{0,1,1}(w)\nonumber\\
&+&\frac{3}{2} (w-1)^2 H_{1,0,1}(w)-\frac{1}{12} \pi ^2 \left(15 w^2+66 w+37\right) (w-1) H_1(w)\nonumber\\
&+&\frac{1}{48} \pi ^2 w \left(16 w^2-13 w+4\right) H_0(w)+\frac{1}{32} \left(400 w^3+199 w^2-192 w-212\right) \zeta (3)\nonumber\\
&-&\frac{1}{16} \pi ^2 \left(16 w^3+27 w^2-76\right) \log (2)+\frac{1}{16} (w-1) \left(177 w^2-106 w-86\right)\nonumber\\
&-&\frac{\left(29 w^2+24 w-1\right) (w-1)^2 H_{1,1}(w)}{4 w}+\frac{\left(22 w^3-99 w^2-63 w-2\right) (w-1) H_{1,0}(w)}{8 w}\nonumber\\
&-&\frac{\left(80 w^4-159 w^3-220 w^2+w+2\right) H_{0,1}(w)}{8 w}-\frac{1}{48} \pi ^2 \left(120 w^3+177 w^2-120 w+119\right)\nonumber\\
&+&\frac{1}{16} \left(8 w^3-385 w^2-196 w-4\right) H_0(w)+\frac{(w-1) \left(34 w^3-449 w^2-175 w-2\right) H_1(w)}{16 w},\nonumber\\
g_A(w) &=&\frac{1}{192} \left(w^2-12\right) \big(24 H_{-1,0,0}(1-w)+24 H_{-1,0,1}(1-w)+14 \pi ^2 H_{-1}(1-w)-3 \zeta (3)\nonumber\\&-&18 \pi ^2 \log (2)\big)
+(w-1) \left(5 w^2+8 w+3\right) \left(-H_{-1,0,1}(w)-\frac{1}{12} \pi ^2 H_{-1}(w)\right)+\frac{X_0}{2}H_{1,0,1}(w)\nonumber\\
&+&\frac{1}{6} \left(16 w^3+27 w^2-24 w-13\right) H_{0,0,1}(w)+\frac{1}{24} \left(38 w^3-117 w^2+42 w+34\right) H_{0,1,0}(w)\nonumber\\
&-&\frac{1}{24} \left(124 w^3-231 w^2+72w+68\right) H_{0,1,1}(w)+\frac{1}{12} (w-1) \left(41 w^2-73 w-22\right)H_{1,1,0}(w)\nonumber\\
&-&\frac{1}{32} \pi ^2 w \left(16 w^2+w-4\right) H_0(w)-\frac{1}{24} \pi ^2 (w-1) \left(57 w^2+45 w-16\right) H_1(w)\nonumber\\
&-&\frac{1}{64} \left(560 w^3-425 w^2-96 w-36\right)\zeta(3)+\frac{1}{32} \pi ^2 \left(16 w^3+27 w^2-76\right) \log (2)\nonumber\\
&+&\frac{1}{72} \left(466 w^2-827 w-485\right) (w-1) H_{1,0}(w)+\frac{1}{72} \left(32 w^3+759 w^2+354 w-224\right) H_{0,1}(w)\nonumber\\
&+&\frac{1}{24} (166 w+65) (w-1)^2 H_{1,1}(w)-\frac{1}{864} \pi ^2 \left(2614 w^3+1005 w^2-1272 w-505\right)\nonumber\\
&+&\frac{1}{288} w \left(2542 w^2-2317 w-2420\right) H_0(w)+\frac{(w-1) \left(1352 w^3-1387 w^2-1873 w+66\right) H_1(w)}{144 w}\nonumber\\
&+&\frac{1}{576} (w-1) \left(1188 w^2-3381 w-785\right).
\eqa

Near the $w=0$ and $w=1$ boundaries, the 
decay width contains logarithmic terms.
Near $w=0$, we find 
\bqa
X_l & = &\ln(w)\bigg(-\frac{1}{3} (2 w+1) (w-1)^2 (H_{0,1}(w)+2H_{1,1}(w))-\frac{1}{18} \left(38 w^3-93 w^2+18 w+37\right) H_1(w)\nonumber\\
& &+\frac{1}{36} w \left(-106 w^2+25 w+86\right)\bigg)+\cdots,
\\
X_F&=&\ln(w)\bigg(\frac{1}{4} \left(w^2-12\right) H_{-1,0}(1-w)+\frac{1}{4} \left(2 w^3-15 w^2+10 w+12\right) H_{0,1}(w)+\frac{1}{2} \left(w^3-26 w^2-w+26\right) H_{1,1}(w)\nonumber\\
& &+\left(2 w^3-w^2+8 w+\frac{3}{2}\right) H_{0,0,1}(w)+\frac{1}{2} \left(-2 w^3+11 w^2+28 w+1\right)
   H_{0,1,1}(w)+\frac{3}{2} (w-1)^2 (2 w+1) H_{1,0,1}(w)\nonumber\\
& &-\frac{(w-1) \left(\left(4 \pi ^2-66\right) w^3+\left(297-2 \pi ^2\right) w^2+\left(189-2 \pi ^2\right) w+6\right) H_1(w)}{24 w}+\frac{1}{12} \pi ^2 \left(4 w^3-3
   w^2+w-3\right)\nonumber\\
& &+\frac{w^3}{2}-\frac{385 w^2}{16}-\frac{49 w}{4}-\frac{1}{4}\bigg)+\cdots,\\
X_A&= &\ln(w)\bigg(-\frac{1}{8} \left(w^2-12\right) H_{-1,0}(1-w)+\frac{1}{24} \left(38 w^3-117 w^2+42 w+34\right) H_{0,1}(w)\nonumber\\
&& +\frac{1}{12} \left(41 w^3-114 w^2+51 w+22\right) H_{1,1}(w)\nonumber\\
& &+\left(2 w^2+w+\frac{1}{4}\right) H_{0,0,1}(w)+\frac{1}{4} \left(2 w^3+13 w^2+12
   w+3\right) H_{0,1,1}(w)-\frac{1}{4} (w-1)^2 (2 w+1) H_{1,0,1}(w)\nonumber\\
& &-\frac{1}{72} (w-1) \left(-466 w^2+9 \pi ^2 \left(2 w^2-w-1\right)+827 w+485\right) H_1(w)-\frac{1}{24} \pi ^2 \left(12 w^3+w^2-3
   w-3\right)\nonumber\\
& &+\frac{1271 w^3}{144}-\frac{2317 w^2}{288}-\frac{605 w}{72}\bigg)+\cdots,
\eqa
where the omitted parts do not contain any logarithms.
$X_h$ does not have $\ln(w)$ terms.

Near $w=1$, we discover the  logarithmic structures, 
\bqa
X_l& = &\frac{\ln ^2(1-w)}{6} (w-1)^2 \left((4 w+2)
   H_0(w)-4 w-5\right) \nonumber \\
&&+
\frac{1}{36} \ln (1-w) \bigg(-12 (2 w+1) (w-1)^2 H_{0,0}(w)+24 (2 w+1) (w-1)^2 H_{1,0}(w)\nonumber\\
& &-4 \left(7 w^3-39 w^2+15 w+5\right) H_0(w)+\frac{\left(124 w^3-35 w^2-143 w+6\right) (w-1)}{w}-12 \pi ^2 (2 w+1) (w-1)^2\bigg)\nonumber\\
&&+\cdots, \\
X_h & = &\frac{1}{54} \ln (1-w) \bigg(-18 \left(2 w^3-3 w^2-12 w+1\right) H_{0,0}(w)+\frac{6 \left(19 w^4+32 w^3-18 w^2-8 w+23\right) H_0(w)}{w-1}\nonumber\\
& & -265 w^3-168 w^2+498 w-\frac{9}{w}-344\bigg)+\cdots,\\
X_F & = & \frac{1}{8} \ln ^2(1-w) \bigg(-\left(w^2-12\right) (H_{-1}(1-w)-\log (2))+4 \left(2 w^2-7 w-16\right) w H_{0,0}(w)\nonumber\\
& &+\left(4 w^3+57 w^2+4 w-54\right) H_0(w)-\frac{(w-1)^2 \left(29 w^2+24 w-1\right)}{w}\bigg)\nonumber\\
& &\frac{1}{48} \ln (1-w) \bigg(-\left(w^2-12\right) \left(-12 H_{0,-1}(1-w)+12 \log (2) H_{0}(1-w)+\pi ^2\right)\nonumber\\
& &-96 \left(5 w^2+8 w+3\right) (w-1) H_{-1,0}(w)+24 \left(18 w^3-9 w^2-4 w+3\right) H_{0,0}(w)\nonumber\\
& &+12 \left(4 w^3+57 w^2+4 w-54\right)H_{1,0}(w)+96 \left(2 w^3-3 w^2+4 w+1\right) H_{0,-1,0}(w)\nonumber\\
& &+24 \left(4 w^3-2 w^2+4 w+3\right) H_{0,0,0}(w)-48 w \left(2 w^2-7 w-16\right) \left(-H_{0,1,0}(w)-H_{1,0,0}(w)\right)\nonumber\\
& &+48 (2 w+1) (w-1)^2 H_{0,1,0}(w)+144 (2 w+1) (w-1)^2H_{1,0,0}(w)-4 \pi ^2 \left(18 w^3-3 w^2+76 w+15\right) H_0(w)\nonumber\\
& &+\frac{6 \left(80 w^4-159 w^3-220 w^2+w+2\right) H_0(w)}{w}+4 \pi ^2 \left(15 w^2+66 w+37\right) (w-1)-72 (w-1)^2 H_{1,0}(w)\nonumber\\
& &-\frac{3 \left(34 w^3-449 w^2-175 w-2\right) (w-1)}{w}-288 (2 w+1) (w-1)^2 \zeta(3)\bigg)+\cdots,\\
X_A & =  & 
\frac{1}{48} \ln ^2(1-w) \bigg(-12 \left(8 w^2+4 w+1\right) H_{0,0}(w)-\left(124 w^3-231 w^2+72 w+68\right) H_0(w)\nonumber\\
& &+3 \left(w^2-12\right) \left(H_{-1}(1-w)-\log (2)\right)+(166 w+65) (w-1)^2\bigg)\nonumber\\
&&+\frac{1}{144} \ln (1-w) \bigg(144 (w-1) \left(5 w^2+8 w+3\right) H_{-1,0}(w)-24 \left(16 w^3+27 w^2-24 w-13\right) H_{0,0}(w)\nonumber\\
& &-6 \left(124 w^3-231 w^2+72 w+68\right) H_{1,0}(w)-144 \left(2 w^3-3 w^2+4 w+1\right) H_{0,-1,0}(w)\nonumber\\
& & +72 \left(8 w^2+4 w+1\right)\left(-H_{0,1,0}(w)-H_{1,0,0}(w)\right)-18 \left(w^2-12\right) \left(H_{0,-1}(1-w)-\log (2) H_0(1-w)-\frac{\pi ^2}{12}\right)\nonumber\\
& &-72 (w-1)^2 (2 w+1)H_{1,0}(w)-144 (w-1)^2 (2 w+1) H_{1,0,0}(w)-6 \pi ^2 \left(10 w^3+33 w^2+44 w+11\right) H_0(w)\nonumber\\
& &-2 \left(32 w^3+759 w^2+354 w-224\right) H_0(w)+6 \pi ^2 (w-1) \left(57 w^2+45 w-16\right)+216 (w-1)^2 (2 w+1) \zeta (3)\nonumber\\
& &+36 \left(8 w^2+16 w+1\right) H_{0,0,0}(w)-\frac{(w-1) \left(1352 w^3-1387 w^2-1873 w+66\right)}{w}\bigg)+\cdots.
\eqa

The top-quark width including off-shell $W$ boson contribution is given by 
\bqa
\Tilde{\Gamma}_t=\frac{\Gamma_0}{\pi}\left[\Tilde{X}_0+\frac{\alpha_s}{\pi} C_F \Tilde{X}_1+\left(\frac{\alpha_s}{\pi}\right)^2\Tilde{X}_2\right]
\eqa
with
\bqa
\Tilde{X}_0=2r w (2w-1)-\frac{1}{2}\left[(2 (r-i) w-i) ((r-i) w+i)^2 G_{w+i r w}(1)+(2 (r+i) w+i) ((r+i) w-i)^2  G_{w-i r w}(1)\right]
\eqa
and
\bqa
\Tilde{X}_1&=&\frac{1}{24} ((r-i) w+i) \left(4 \pi ^2 \left(2 (r-i)^2 w^2+i r w+w+1\right)-3 \left(6 (r-i)^2 w^2+(9+9 i r) w+5\right)\right) G_{w+i r
   w}(1)\nonumber\\
  & +&\frac{1}{24} ((r+i) w-i) \left(4 \pi ^2 \left(2 (r+i)^2 w^2-i r w+w+1\right)-3 \left(6 (r+i)^2 w^2+(9-9 i r) w+5\right)\right) G_{w-i r
   w}(1)\nonumber\\
   &-&\frac{1}{2} (r+i) w \left(2 (r+i)^2 w^2+(-1+i r) w+1\right) G_{w-i r w,0}(1)\nonumber\\
   &-&\frac{1}{2} (r-i) w \left(2 (r-i)^2 w^2+(-1-i r) w+1\right) G_{w+i r w,0}(1)\nonumber\\
   &+&\frac{1}{4} (4 (r-i) w-5 i) ((r-i) w+i)^2 G_{w+i r w,1}(1)+\frac{1}{4} (4 (r+i) w+5 i)((r+i) w-i)^2  G_{w-i r w,1}(1)\nonumber\\
   &+&\frac{1}{2} (2 (r-i) w-i) ((r-i) w+i)^2 G_{w+i r w,1,0}(1)+\frac{1}{2} (2 (r+i) w+i) ((r+i) w-i)^2  G_{w-i r w,1,0}(1)\nonumber\\
   &-&\frac{1}{2}   (2 (r+i) w+i)((r+i) w-i)^2  G_{w-i r w,0,1}(1)-\frac{1}{2} (2 (r-i) w-i) ((r-i) w+i)^2 G_{w+i r w,0,1}(1)\nonumber\\
   &+&\pi ^2 r (1-2 w) w+r w (3 w-5).
\eqa
Here $r=\frac{\Gamma_W}{m_W}$ and $G_{a_1,a_2,...,a_n}(x)$ are multiple polylogarithms defined in \cite{Goncharov:1998kja}. The coefficient $\Tilde{X}_2$ can be found in the ancillary file associated with the arXiv submission  of this article, arXiv:2212.06341.

\bibliography{loop}

\providecommand{\href}[2]{#2}\begingroup\raggedright\begin{thebibliography}{10}

\bibitem{D0:1995jca}
{\scshape D0} collaboration, S.~Abachi et~al., \emph{{Observation of the top
  quark}}, \href{https://doi.org/10.1103/PhysRevLett.74.2632}{\emph{Phys. Rev.
  Lett.} {\bfseries 74} (1995) 2632--2637},
  [\href{https://arxiv.org/abs/hep-ex/9503003}{{\ttfamily hep-ex/9503003}}].

\bibitem{CDF:1995wbb}
{\scshape CDF} collaboration, F.~Abe et~al., \emph{{Observation of top quark
  production in $\bar{p}p$ collisions}},
  \href{https://doi.org/10.1103/PhysRevLett.74.2626}{\emph{Phys. Rev. Lett.}
  {\bfseries 74} (1995) 2626--2631},
  [\href{https://arxiv.org/abs/hep-ex/9503002}{{\ttfamily hep-ex/9503002}}].

\bibitem{Bigi:1986jk}
I.~I.~Y. Bigi, Y.~L. Dokshitzer, V.~A. Khoze, J.~H. Kuhn and P.~M. Zerwas,
  \emph{{Production and Decay Properties of Ultraheavy Quarks}},
  \href{https://doi.org/10.1016/0370-2693(86)91275-X}{\emph{Phys. Lett. B}
  {\bfseries 181} (1986) 157--163}.

\bibitem{ATLAS:2019onj}
{\scshape ATLAS} collaboration, \emph{{Measurement of the top-quark decay width
  in top-quark pair events in the dilepton channel at $\sqrt{s}=13$ TeV with
  the ATLAS detector}}, .

\bibitem{CMS:2014mxl}
{\scshape CMS} collaboration, V.~Khachatryan et~al., \emph{{Measurement of the
  ratio $\mathcal B(t \to Wb)/\mathcal B(t \to Wq)$ in pp collisions at
  $\sqrt{s}$ = 8 TeV}},
  \href{https://doi.org/10.1016/j.physletb.2014.06.076}{\emph{Phys. Lett. B}
  {\bfseries 736} (2014) 33--57},
  [\href{https://arxiv.org/abs/1404.2292}{{\ttfamily 1404.2292}}].

\bibitem{Giardino:2017hva}
P.~P. Giardino and C.~Zhang, \emph{{Probing the top-quark width using the
  charge identification of $b$ jets}},
  \href{https://doi.org/10.1103/PhysRevD.96.011901}{\emph{Phys. Rev. D}
  {\bfseries 96} (2017) 011901},
  [\href{https://arxiv.org/abs/1702.06996}{{\ttfamily 1702.06996}}].

\bibitem{Baskakov:2018huw}
A.~Baskakov, E.~Boos and L.~Dudko, \emph{{Model independent top quark width
  measurement using a combination of resonant and nonresonant cross sections}},
  \href{https://doi.org/10.1103/PhysRevD.98.116011}{\emph{Phys. Rev. D}
  {\bfseries 98} (2018) 116011},
  [\href{https://arxiv.org/abs/1807.11193}{{\ttfamily 1807.11193}}].

\bibitem{Herwig:2019obz}
C.~Herwig, T.~Je\v{z}o and B.~Nachman, \emph{{Extracting the Top-Quark Width
  from Nonresonant Production}},
  \href{https://doi.org/10.1103/PhysRevLett.122.231803}{\emph{Phys. Rev. Lett.}
  {\bfseries 122} (2019) 231803},
  [\href{https://arxiv.org/abs/1903.10519}{{\ttfamily 1903.10519}}].

\bibitem{Martinez:2002st}
M.~Martinez and R.~Miquel, \emph{{Multiparameter fits to the t anti-t threshold
  observables at a future e+ e- linear collider}},
  \href{https://doi.org/10.1140/epjc/s2002-01094-1}{\emph{Eur. Phys. J. C}
  {\bfseries 27} (2003) 49--55},
  [\href{https://arxiv.org/abs/hep-ph/0207315}{{\ttfamily hep-ph/0207315}}].

\bibitem{Horiguchi:2013wra}
T.~Horiguchi, A.~Ishikawa, T.~Suehara, K.~Fujii, Y.~Sumino, Y.~Kiyo et~al.,
  \emph{{Study of top quark pair production near threshold at the ILC}},
  \href{https://arxiv.org/abs/1310.0563}{{\ttfamily 1310.0563}}.

\bibitem{Li:2022iav}
Z.~Li, X.~Sun, Y.~Fang, G.~Li, S.~Xin, S.~Wang et~al., \emph{{Top quark mass
  measurements at the $t\bar{t}$ threshold with CEPC}},
  \href{https://arxiv.org/abs/2207.12177}{{\ttfamily 2207.12177}}.

\bibitem{CLICdp:2018esa}
{\scshape CLICdp} collaboration, H.~Abramowicz et~al., \emph{{Top-Quark Physics
  at the CLIC Electron-Positron Linear Collider}},
  \href{https://doi.org/10.1007/JHEP11(2019)003}{\emph{JHEP} {\bfseries 11}
  (2019) 003}, [\href{https://arxiv.org/abs/1807.02441}{{\ttfamily
  1807.02441}}].

\bibitem{Jezabek:1988iv}
M.~Jezabek and J.~H. Kuhn, \emph{{QCD Corrections to Semileptonic Decays of
  Heavy Quarks}},
  \href{https://doi.org/10.1016/0550-3213(89)90108-9}{\emph{Nucl. Phys. B}
  {\bfseries 314} (1989) 1--6}.

\bibitem{Czarnecki:1990kv}
A.~Czarnecki, \emph{{QCD corrections to the decay t ---\ensuremath{>} W b in
  dimensional regularization}},
  \href{https://doi.org/10.1016/0370-2693(90)90571-M}{\emph{Phys. Lett. B}
  {\bfseries 252} (1990) 467--470}.

\bibitem{Li:1990qf}
C.~S. Li, R.~J. Oakes and T.~C. Yuan, \emph{{QCD corrections to $t \to W^{+}
  b$}}, \href{https://doi.org/10.1103/PhysRevD.43.3759}{\emph{Phys. Rev. D}
  {\bfseries 43} (1991) 3759--3762}.

\bibitem{Denner:1990ns}
A.~Denner and T.~Sack, \emph{{The Top width}},
  \href{https://doi.org/10.1016/0550-3213(91)90530-B}{\emph{Nucl. Phys. B}
  {\bfseries 358} (1991) 46--58}.

\bibitem{Eilam:1991iz}
G.~Eilam, R.~R. Mendel, R.~Migneron and A.~Soni, \emph{{Radiative corrections
  to top quark decay}},
  \href{https://doi.org/10.1103/PhysRevLett.66.3105}{\emph{Phys. Rev. Lett.}
  {\bfseries 66} (1991) 3105--3108}.

\bibitem{Czarnecki:1998qc}
A.~Czarnecki and K.~Melnikov, \emph{{Two loop QCD corrections to top quark
  width}}, \href{https://doi.org/10.1016/S0550-3213(98)00844-X}{\emph{Nucl.
  Phys. B} {\bfseries 544} (1999) 520--531},
  [\href{https://arxiv.org/abs/hep-ph/9806244}{{\ttfamily hep-ph/9806244}}].

\bibitem{Chetyrkin:1999ju}
K.~G. Chetyrkin, R.~Harlander, T.~Seidensticker and M.~Steinhauser,
  \emph{{Second order QCD corrections to Gamma(t ---\ensuremath{>} W b)}},
  \href{https://doi.org/10.1103/PhysRevD.60.114015}{\emph{Phys. Rev. D}
  {\bfseries 60} (1999) 114015},
  [\href{https://arxiv.org/abs/hep-ph/9906273}{{\ttfamily hep-ph/9906273}}].

\bibitem{Blokland:2004ye}
I.~R. Blokland, A.~Czarnecki, M.~Slusarczyk and F.~Tkachov, \emph{{Heavy to
  light decays with a two loop accuracy}},
  \href{https://doi.org/10.1103/PhysRevLett.93.062001}{\emph{Phys. Rev. Lett.}
  {\bfseries 93} (2004) 062001},
  [\href{https://arxiv.org/abs/hep-ph/0403221}{{\ttfamily hep-ph/0403221}}].

\bibitem{Blokland:2005vq}
I.~R. Blokland, A.~Czarnecki, M.~Slusarczyk and F.~Tkachov,
  \emph{{Next-to-next-to-leading order calculations for heavy-to-light
  decays}}, \href{https://doi.org/10.1103/PhysRevD.71.054004}{\emph{Phys. Rev.
  D} {\bfseries 71} (2005) 054004},
  [\href{https://arxiv.org/abs/hep-ph/0503039}{{\ttfamily hep-ph/0503039}}].

\bibitem{Czarnecki:2001cz}
A.~Czarnecki and K.~Melnikov, \emph{{Semileptonic b ---\ensuremath{>} u decays:
  Lepton invariant mass spectrum}},
  \href{https://doi.org/10.1103/PhysRevLett.88.131801}{\emph{Phys. Rev. Lett.}
  {\bfseries 88} (2002) 131801},
  [\href{https://arxiv.org/abs/hep-ph/0112264}{{\ttfamily hep-ph/0112264}}].

\bibitem{Gao:2012ja}
J.~Gao, C.~S. Li and H.~X. Zhu, \emph{{Top Quark Decay at Next-to-Next-to
  Leading Order in QCD}},
  \href{https://doi.org/10.1103/PhysRevLett.110.042001}{\emph{Phys. Rev. Lett.}
  {\bfseries 110} (2013) 042001},
  [\href{https://arxiv.org/abs/1210.2808}{{\ttfamily 1210.2808}}].

\bibitem{Brucherseifer:2013iv}
M.~Brucherseifer, F.~Caola and K.~Melnikov, \emph{{$\mathcal O(\alpha_s^2)$
  corrections to fully-differential top quark decays}},
  \href{https://doi.org/10.1007/JHEP04(2013)059}{\emph{JHEP} {\bfseries 04}
  (2013) 059}, [\href{https://arxiv.org/abs/1301.7133}{{\ttfamily 1301.7133}}].

\bibitem{Czarnecki:2010gb}
A.~Czarnecki, J.~G. Korner and J.~H. Piclum, \emph{{Helicity fractions of W
  bosons from top quark decays at NNLO in QCD}},
  \href{https://doi.org/10.1103/PhysRevD.81.111503}{\emph{Phys. Rev. D}
  {\bfseries 81} (2010) 111503},
  [\href{https://arxiv.org/abs/1005.2625}{{\ttfamily 1005.2625}}].

\bibitem{Czarnecki:2018vwh}
A.~Czarnecki, S.~Groote, J.~G. K\"orner and J.~H. Piclum, \emph{{NNLO QCD
  corrections to the polarized top quark decay $t(\uparrow) \to X_b+W^+$}},
  \href{https://doi.org/10.1103/PhysRevD.97.094008}{\emph{Phys. Rev. D}
  {\bfseries 97} (2018) 094008},
  [\href{https://arxiv.org/abs/1803.03658}{{\ttfamily 1803.03658}}].

\bibitem{Meng:2022htg}
R.-Q. Meng, S.-Q. Wang, T.~Sun, C.-Q. Luo, J.-M. Shen and X.-G. Wu, \emph{{QCD
  improved top-quark decay at next-to-next-to-leading order}},
  \href{https://arxiv.org/abs/2202.09978}{{\ttfamily 2202.09978}}.

\bibitem{Hahn:2000kx}
T.~Hahn, \emph{{Generating Feynman diagrams and amplitudes with FeynArts 3}},
  \href{https://doi.org/10.1016/S0010-4655(01)00290-9}{\emph{Comput. Phys.
  Commun.} {\bfseries 140} (2001) 418--431},
  [\href{https://arxiv.org/abs/hep-ph/0012260}{{\ttfamily hep-ph/0012260}}].

\bibitem{Shtabovenko:2020gxv}
V.~Shtabovenko, R.~Mertig and F.~Orellana, \emph{{FeynCalc 9.3: New features
  and improvements}},
  \href{https://doi.org/10.1016/j.cpc.2020.107478}{\emph{Comput. Phys. Commun.}
  {\bfseries 256} (2020) 107478},
  [\href{https://arxiv.org/abs/2001.04407}{{\ttfamily 2001.04407}}].

\bibitem{Tkachov:1981wb}
F.~V. Tkachov, \emph{{A Theorem on Analytical Calculability of Four Loop
  Renormalization Group Functions}},
  \href{https://doi.org/10.1016/0370-2693(81)90288-4}{\emph{Phys. Lett. B}
  {\bfseries 100} (1981) 65--68}.

\bibitem{Chetyrkin:1981qh}
K.~G. Chetyrkin and F.~V. Tkachov, \emph{{Integration by Parts: The Algorithm
  to Calculate beta Functions in 4 Loops}},
  \href{https://doi.org/10.1016/0550-3213(81)90199-1}{\emph{Nucl. Phys. B}
  {\bfseries 192} (1981) 159--204}.

\bibitem{Smirnov:2019qkx}
A.~V. Smirnov and F.~S. Chuharev, \emph{{FIRE6: Feynman Integral REduction with
  Modular Arithmetic}},
  \href{https://doi.org/10.1016/j.cpc.2019.106877}{\emph{Comput. Phys. Commun.}
  {\bfseries 247} (2020) 106877},
  [\href{https://arxiv.org/abs/1901.07808}{{\ttfamily 1901.07808}}].

\bibitem{doi:10.1063/1.1703676}
R.~E. Cutkosky, \emph{Singularities and discontinuities of feynman amplitudes},
  \href{https://doi.org/10.1063/1.1703676}{\emph{Journal of Mathematical
  Physics} {\bfseries 1} (1960) 429--433},
  [\href{https://arxiv.org/abs/https://doi.org/10.1063/1.1703676}{{\ttfamily
  https://doi.org/10.1063/1.1703676}}].

\bibitem{Kotikov:1990kg}
A.~V. Kotikov, \emph{{Differential equations method: New technique for massive
  Feynman diagrams calculation}},
  \href{https://doi.org/10.1016/0370-2693(91)90413-K}{\emph{Phys. Lett.}
  {\bfseries B254} (1991) 158--164}.

\bibitem{Kotikov:1991pm}
A.~V. Kotikov, \emph{{Differential equation method: The Calculation of N point
  Feynman diagrams}}, \href{https://doi.org/10.1016/0370-2693(91)90536-Y,
  10.1016/0370-2693(92)91582-T}{\emph{Phys. Lett.} {\bfseries B267} (1991)
  123--127}.

\bibitem{Henn:2013pwa}
J.~M. Henn, \emph{{Multiloop integrals in dimensional regularization made
  simple}}, \href{https://doi.org/10.1103/PhysRevLett.110.251601}{\emph{Phys.
  Rev. Lett.} {\bfseries 110} (2013) 251601},
  [\href{https://arxiv.org/abs/1304.1806}{{\ttfamily 1304.1806}}].

\bibitem{vanRitbergen:1999fi}
T.~van Ritbergen and R.~G. Stuart, \emph{{On the precise determination of the
  Fermi coupling constant from the muon lifetime}},
  \href{https://doi.org/10.1016/S0550-3213(99)00572-6}{\emph{Nucl. Phys. B}
  {\bfseries 564} (2000) 343--390},
  [\href{https://arxiv.org/abs/hep-ph/9904240}{{\ttfamily hep-ph/9904240}}].

\bibitem{Liu:2017jxz}
X.~Liu, Y.-Q. Ma and C.-Y. Wang, \emph{{A Systematic and Efficient Method to
  Compute Multi-loop Master Integrals}},
  \href{https://doi.org/10.1016/j.physletb.2018.02.026}{\emph{Phys. Lett. B}
  {\bfseries 779} (2018) 353--357},
  [\href{https://arxiv.org/abs/1711.09572}{{\ttfamily 1711.09572}}].

\bibitem{Liu:2022chg}
X.~Liu and Y.-Q. Ma, \emph{{AMFlow: A Mathematica package for Feynman integrals
  computation via auxiliary mass flow}},
  \href{https://doi.org/10.1016/j.cpc.2022.108565}{\emph{Comput. Phys. Commun.}
  {\bfseries 283} (2023) 108565},
  [\href{https://arxiv.org/abs/2201.11669}{{\ttfamily 2201.11669}}].

\bibitem{Ferguson1992}
H.~Ferguson and D.~Bailey, \emph{{A polynomial time, numerically stable integer
  relation algorithm}},  {1992}.

\bibitem{Ferguson:1999aa}
H.~Ferguson, D.~Beiley and S.~Arno, \emph{{Analysis of PSLQ, an integer
  relation finding algorithm}}, {\emph{Math. Comp.} {\bfseries 68} (1999) 351}.

\bibitem{Goncharov:1998kja}
A.~B. Goncharov, \emph{{Multiple polylogarithms, cyclotomy and modular
  complexes}}, \href{https://doi.org/10.4310/MRL.1998.v5.n4.a7}{\emph{Math.
  Res. Lett.} {\bfseries 5} (1998) 497--516},
  [\href{https://arxiv.org/abs/1105.2076}{{\ttfamily 1105.2076}}].

\bibitem{Remiddi:1999ew}
E.~Remiddi and J.~A.~M. Vermaseren, \emph{{Harmonic polylogarithms}},
  \href{https://doi.org/10.1142/S0217751X00000367}{\emph{Int. J. Mod. Phys. A}
  {\bfseries 15} (2000) 725--754},
  [\href{https://arxiv.org/abs/hep-ph/9905237}{{\ttfamily hep-ph/9905237}}].

\bibitem{vanRitbergen:1999gs}
T.~van Ritbergen, \emph{{The Second order QCD contribution to the semileptonic
  b ---\ensuremath{>} u decay rate}},
  \href{https://doi.org/10.1016/S0370-2693(99)00407-4}{\emph{Phys. Lett. B}
  {\bfseries 454} (1999) 353--358},
  [\href{https://arxiv.org/abs/hep-ph/9903226}{{\ttfamily hep-ph/9903226}}].

\bibitem{vanRitbergen:1998yd}
T.~van Ritbergen and R.~G. Stuart, \emph{{Complete two loop quantum
  electrodynamic contributions to the muon lifetime in the Fermi model}},
  \href{https://doi.org/10.1103/PhysRevLett.82.488}{\emph{Phys. Rev. Lett.}
  {\bfseries 82} (1999) 488--491},
  [\href{https://arxiv.org/abs/hep-ph/9808283}{{\ttfamily hep-ph/9808283}}].

\bibitem{MuLan:2012sih}
{\scshape MuLan} collaboration, V.~Tishchenko et~al., \emph{{Detailed Report of
  the MuLan Measurement of the Positive Muon Lifetime and Determination of the
  Fermi Constant}},
  \href{https://doi.org/10.1103/PhysRevD.87.052003}{\emph{Phys. Rev. D}
  {\bfseries 87} (2013) 052003},
  [\href{https://arxiv.org/abs/1211.0960}{{\ttfamily 1211.0960}}].

\bibitem{ParticleDataGroup:2022pth}
{\scshape Particle Data Group} collaboration, R.~L. Workman et~al.,
  \emph{{Review of Particle Physics}},
  \href{https://doi.org/10.1093/ptep/ptac097}{\emph{PTEP} {\bfseries 2022}
  (2022) 083C01}.

\bibitem{Bohm:1986rj}
M.~Bohm, H.~Spiesberger and W.~Hollik, \emph{{On the One Loop Renormalization
  of the Electroweak Standard Model and Its Application to Leptonic
  Processes}}, \href{https://doi.org/10.1002/prop.19860341102}{\emph{Fortsch.
  Phys.} {\bfseries 34} (1986) 687--751}.

\bibitem{Denner:1990tx}
A.~Denner and T.~Sack, \emph{{The W boson width}},
  \href{https://doi.org/10.1007/BF01560267}{\emph{Z. Phys. C} {\bfseries 46}
  (1990) 653--663}.

\bibitem{Gardi:1998qr}
E.~Gardi, G.~Grunberg and M.~Karliner, \emph{{Can the QCD running coupling have
  a causal analyticity structure?}},
  \href{https://doi.org/10.1088/1126-6708/1998/07/007}{\emph{JHEP} {\bfseries
  07} (1998) 007}, [\href{https://arxiv.org/abs/hep-ph/9806462}{{\ttfamily
  hep-ph/9806462}}].

\bibitem{Deur:2016tte}
A.~Deur, S.~J. Brodsky and G.~F. de~Teramond, \emph{{The QCD Running
  Coupling}}, \href{https://doi.org/10.1016/j.ppnp.2016.04.003}{\emph{Nucl.
  Phys.} {\bfseries 90} (2016) 1},
  [\href{https://arxiv.org/abs/1604.08082}{{\ttfamily 1604.08082}}].

\bibitem{Beneke:2022obx}
M.~Beneke, M.~Garny, S.~Jaskiewicz, J.~Strohm, R.~Szafron, L.~Vernazza et~al.,
  \emph{{Next-to-leading power endpoint factorization and resummation for
  off-diagonal \textquotedblleft{}gluon\textquotedblright{} thrust}},
  \href{https://doi.org/10.1007/JHEP07(2022)144}{\emph{JHEP} {\bfseries 07}
  (2022) 144}, [\href{https://arxiv.org/abs/2205.04479}{{\ttfamily
  2205.04479}}].

\bibitem{Beneke:2021lkq}
M.~Beneke, \emph{{Pole mass renormalon and its ramifications}},
  \href{https://doi.org/10.1140/epjs/s11734-021-00268-w}{\emph{Eur. Phys. J.
  ST} {\bfseries 230} (2021) 2565--2579},
  [\href{https://arxiv.org/abs/2108.04861}{{\ttfamily 2108.04861}}].

\end{thebibliography}\endgroup
\bibliographystyle{JHEP}

\end{document}